\documentclass[preprint,showpacs,pre]{revtex4}

\usepackage{graphicx}
\usepackage{dcolumn}
\usepackage{bm}

\usepackage{graphics}
\begin{document}

\title{Phase diagram of solution of two oppositely
charged polyelectrolytes}

\author{Rui Zhang}
\author{B. I. Shklovskii}
\affiliation{Theoretical Physics Institute, University of
Minnesota, Minneapolis, Minnesota 55455}

\date{\today}

\begin{abstract}
We study a salty water solution of long polyanions (PA) with shorter
polycations (PC) and focus on the role of their Coulomb interaction. A
good example is a solution of DNA and PC which are widely studied
for gene therapy. In such a solution, each PA attracts many PCs to
form a linear complex. When the ratio of total charges of PAs and PCs in
the solution $x$ is close to 1, complexes are neutral and, therefore, they
condense in a macroscopic drop. When $|x|$ is far away from 1,
complexes are strongly charged, negatively at $x < 1$ and positively at $x > 1$. 
Their strong Coulomb repulsion makes the solution of free complexes stable. 
At intermediate $|x|$ some complexes become neutral and, therefore, 
condense in a macroscopic drop, while other complexes become even stronger 
charged and stay free. This phenomenon is called intercomplex disproportionation. 
We calculate a phase diagram of a PA-PC solution in a plane of $x$ and inverse 
screening radius of a monovalent salt. With increasing screening radius $r_s$ 
all phases containing neutral drops shrink. At very large $r_s > r_c$ another, 
intracomplex disproportionation takes place. Namely, one side of each PC complex becomes
neutral and condenses in a small droplet, while the rest of the complex 
forms a linear strongly charged tail. Thus, at very large $r_s$ our phase diagram 
contains a number of new phases with "tadpoles".

\end{abstract}


\maketitle

\section{Introduction}\label{sec:introduction}
Condensation in solution of two oppositely charged
polyelectrolytes (PE) is an important phenomenon in biology and
chemical engineering. One of the most interesting applications is
DNA condensation by polycations (PC), which is widely used in gene
therapy research. A good example is condensation of DNA with
poly-lysine~\cite{Budker}. Complexation of DNA with PCs can invert
the charge of bare DNA and help DNA to penetrate negatively
charged cell membrane. At the same time, adsorbed PC in complexes
or their condensate may protect DNA from digestion by enzymes
inside the cell~\cite{Kabanov}. Tremendous amount of experimental
works have been done in this area~\cite{Kabanov,Budker,van
Zanten,Minagawa,Dunlap,Zinchenko,Bloomfield}.

\begin{figure}[ht]
\begin{center}
\includegraphics[width=0.7\textwidth]{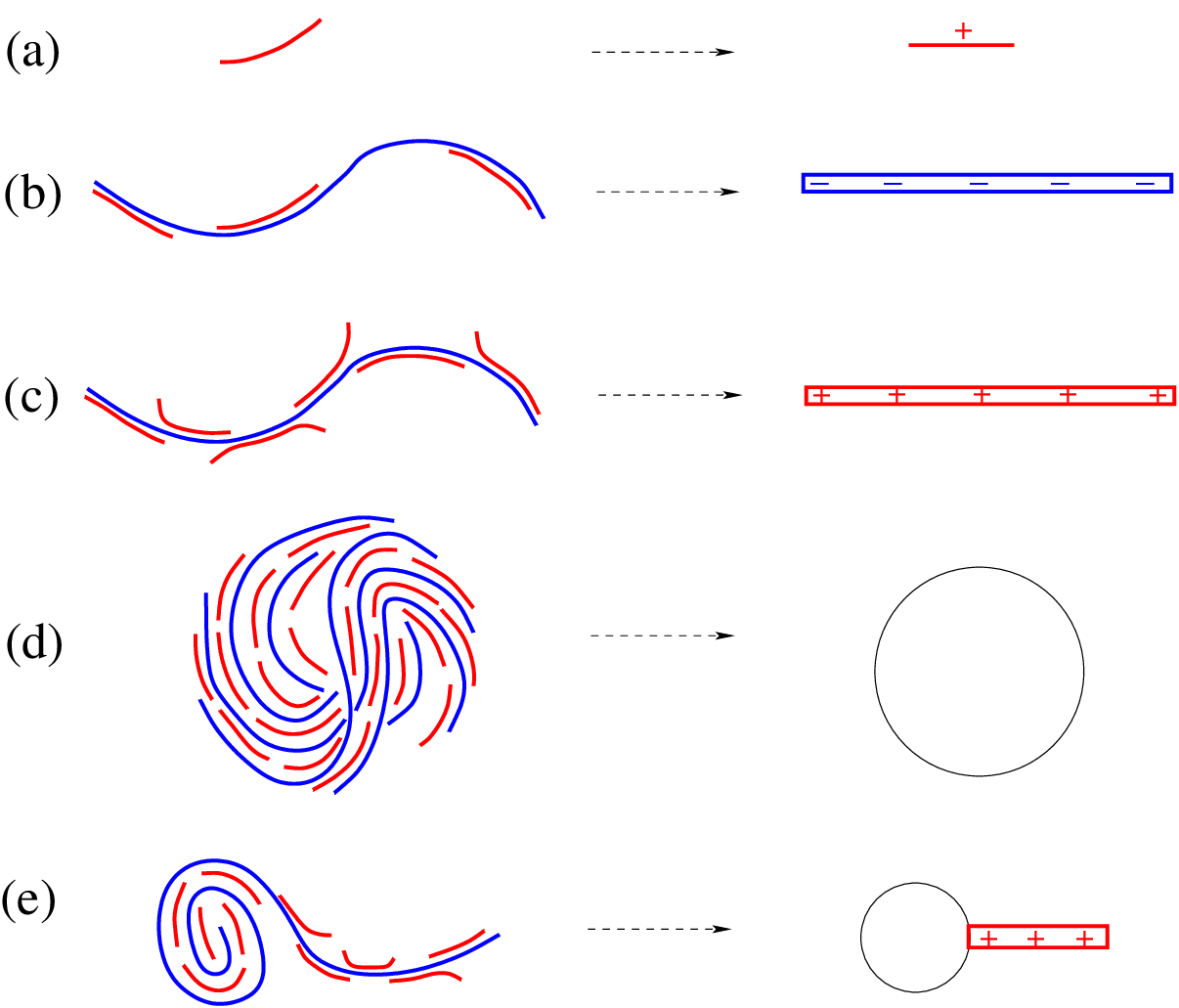}
\end{center}
\caption{Objects in a solution of PA and PC (left
column) and their symbols used in Figs.~\ref{fig:phase},
~\ref{fig:condensate},~\ref{fig:phasehybrid} and~\ref{fig:entropy}
(right column). The long blue polymer is PA and the short red polymer is
PC. (a) a single PC. (b) negative PA-PCs complex. (c) positive
PA-PCs complex. (d) condensate of almost neutral complexes. (e)
tadpole made of one PA-PCs complex. Here only the case of positive
tail is shown. The tail can be negative, too.}\label{fig:PE}
\end{figure}

\begin{figure}[ht]
\begin{center}
\includegraphics[width=1\textwidth]{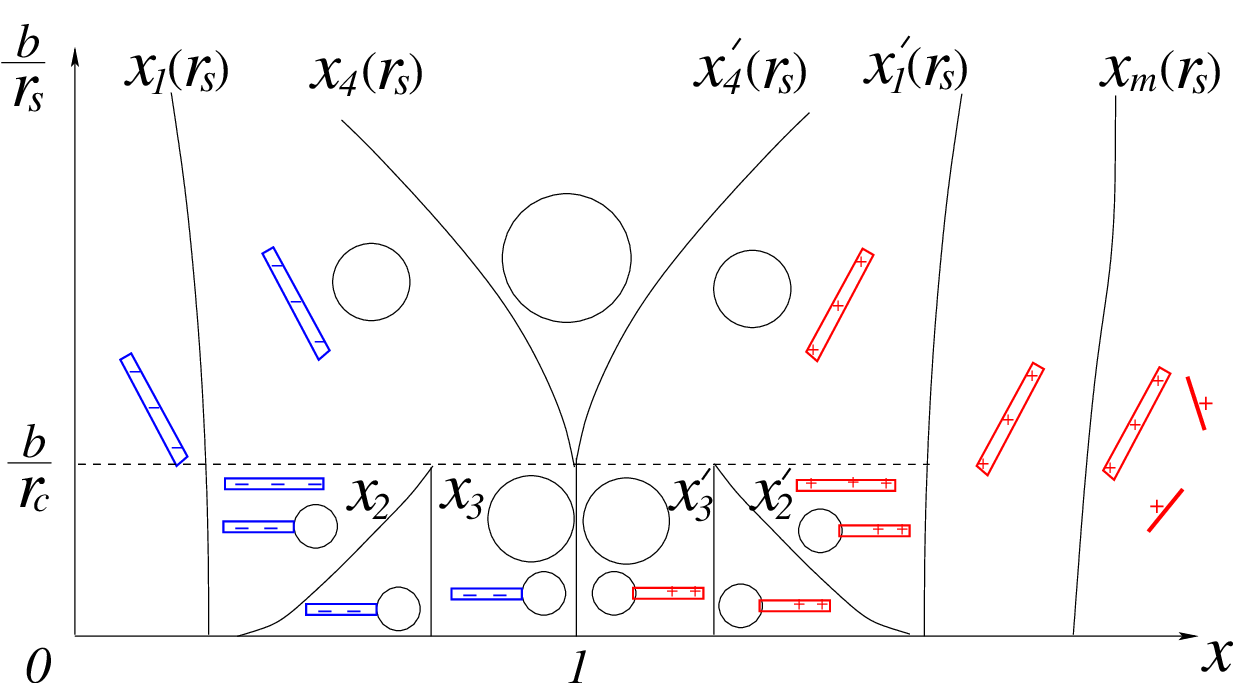}
\end{center}
\caption{Typical phase diagram of a solution of PA and PC. The
horizontal axis $x$ is the ratio of total charges of PCs and PAs in
the solution. The vertical axis shows the ratio  $b/r_s < 1$ of the length
of a monomer $b$ of the PA molecule to the Debye-H\"{u}ckel
screening radius of the salt $r_s$. Given by Eq. (37) length $r_c$ 
is the critical value of $r_s$ at which tadpoles emerge. 
Symbols are explained in Fig.~\ref{fig:PE}.}\label{fig:phase}
\end{figure}

In this paper, motivated by DNA-PC condensation, we study the
equilibrium state of a solution of polyanions (PA) and polycations
(PC) in the presence of monovalent salt and focus on the role of
Coulomb interaction. We assume that both PA and PC are so long
that at room temperature $T$ their translational entropy can be
ignored in comparison with their Coulomb energy. We are
particularly interested in the case when PA is much longer than PC
and many PCs are needed to neutralize one PA. In Fig.~\ref{fig:PE}
we list objects which can appear in such a solution. Each PA
attracts many PCs to form a PA-PCs complex (Fig.~\ref{fig:PE}b,c).
Neutral complexes can further condense in a liquid drop
(Fig.~\ref{fig:PE}d). One complex can form a neutral head and a
charged tail to become a tadpole (Fig.~\ref{fig:PE}e). And it is
possible to have excessive free PCs (Fig.~\ref{fig:PE}a). When and
where these objects exist or co-exist with each other depends on
two dimensionless parameters: the ratio of absolute values of 
total charges of PCs and PAs in the solution, $x$, and the ration $b/r_s$, where $b$ is the size of the
monomer and $r_s$ is the Debye-H\"{u}ckel screening radius
provided by a monovalent salt. The main result of this paper is the
phase diagram in a plane of $x$ and $b/r_s$ shown in
Fig.~\ref{fig:phase}. We present a first theory of broadening 
of the phase of a single drop with decreasing $r_s$ 
(curves $x_4(r_s)$ and $x_4^{\prime}(r_s)$) in Fig.~\ref{fig:phase}). 
We also discover new exotic phases of ``tadpoles".

Up to now, there was no complete theory of phase diagram for such
systems. Previously, the complexation of oppositely charged
polyelectrolytes was studied in a symmetric system in which the
length, concentration and linear charge density of PA and PC are
the same~\cite{Borue,Joanny}. It was shown that even in the
absence of monovalent salt, strongly charged PA and PC form a
single macroscopic drop of neutral dense liquid, which separates
from water. It corresponds to the phase at $x=1$ in our phase
diagram (Fig.~\ref{fig:phase}). On the other hand, the phase
diagram of a solution of DNA and short polyamines was studied in
Refs.~\cite{Olvera,Olvera2,Rouzina} in which translational entropy
of polyamines plays a very important role. We postpone discussion
of the role of translational entropy till the end of this paper
(Sec.~\ref{sec:entropy}).

In our previous works~\cite{Nguyen-reentrant,Rui}, phase diagrams have been
discussed for other systems. In Ref.~\cite{Nguyen-reentrant},
solution of very long s with positive spheres was considered.
This system is similar to chromatin in the sense that each PA
binds many spheres making a long necklace. We also discussed the
phase diagram of a system of oppositely charged spheres in
strongly asymmetric case when each say negative sphere complexes
with many positive ones~\cite{Rui}. Many features of the phase
diagram in Fig.~\ref{fig:phase} are also applicable to these
systems and we will return to them in the conclusion.

Our recent paper studied the exactly same topic~\cite{Rui3} as here.
But it missed two phase coexistence tadpole regions at
$r_s > r_c$ (compare Fig.~2 of Ref.~\cite{Rui3} with
Fig.~\ref{fig:phase} in this paper). The main purpose of the
present paper is correct $r_s > r_c$  part of the diagram. 

It is enough to look at $x<1$ side of Fig.~\ref{fig:phase} to see
the main feature of the phase diagram. It contains three main phases:
the phase of a single drop for $x>x_4$, the phase of free negative
complexes for $x<x_1$, and the `` negative tadpole" phase for $x_2<x<x_3$.
Between these phases there are three regions of phase
coexistence. Namely, they are phases of macroscopic drop with negative complexes,
of negative tadpoles and negative complexes and of negative tadpoles.
The fact that in the last case phases of complexes and tadpoles are mixed  
plays no role in our theory ignoring entropy of mixture.

This phase diagram is similar to the phase
diagram of water in temperature and volume
coordinates~\cite{Landau}, when $r_s/b$ is considered as
temperature and $x$ as volume. The three main phases mentioned above
are like gas, solid and liquid respectively. Essentially, this
analogy originates from the Gibbs' phase rule~\cite{Landau}, which
is crucial to determine our phase diagram (see
Sec.~\ref{sec:screen}). 

Let us now try to understand the physics of this phase diagram. We
start from the horizontal axis ($r_s\rightarrow\infty$) and first
focus on $x<1$ side. In the solution, each PA adsorbs many PCs to
form a complex. When $x\ll 1$, the number of PC is not enough to
neutralize all PAs and each PA-PCs complex is strongly negatively
charged (Fig.~\ref{fig:PE}b). The Coulomb repulsion between
complexes is huge and all complexes stay free, or in other words,
colloidal solution of complexes is stable (see ranges $0<x<x_1$ in
Fig.~\ref{fig:phase}).

\begin{figure}[ht]
\begin{center}
\includegraphics[width=0.5\textwidth]{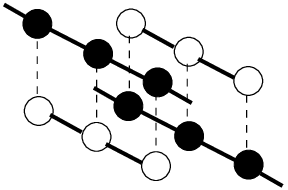}
\end{center}
\caption{A schematic illustration of short-range attraction
between neutral complexes in a condensed liquid drop using the
model of strongly charged PA and PC. A portion of two complexes in
the liquid drop is shown. PA and PC charges are shown by black and
white dots correspondingly. The dashed lines show two complexes
sitting in parallel planes. The complexes attract each other
because charges of the same sign are farther away than charges of
the opposite sign. }\label{fig:correlation}
\end{figure}

When $x=1$, each PA-PCs complex is neutral and there is no Coulomb
repulsion between them. They all condense to form a macroscopic
strong correlated liquid drop (see Fig.~\ref{fig:phase}). Due to
the orderly arrangement of positive and negative charges in the
drop, a certain amount of short-range correlation energy is gained
(Fig.~\ref{fig:correlation}). We define $\varepsilon<0$ as the
energy gain of a neutral complex in the macroscopic drop. Note
that monomers on the surface of the drop can not gain as much
energy as monomers inside. This defines the surface energy of the
drop which plays an important role in the competition between two
kinds of disproportionations (see the next paragraph).

\begin{figure}[ht]
\begin{center}
\includegraphics[width=0.6\textwidth]{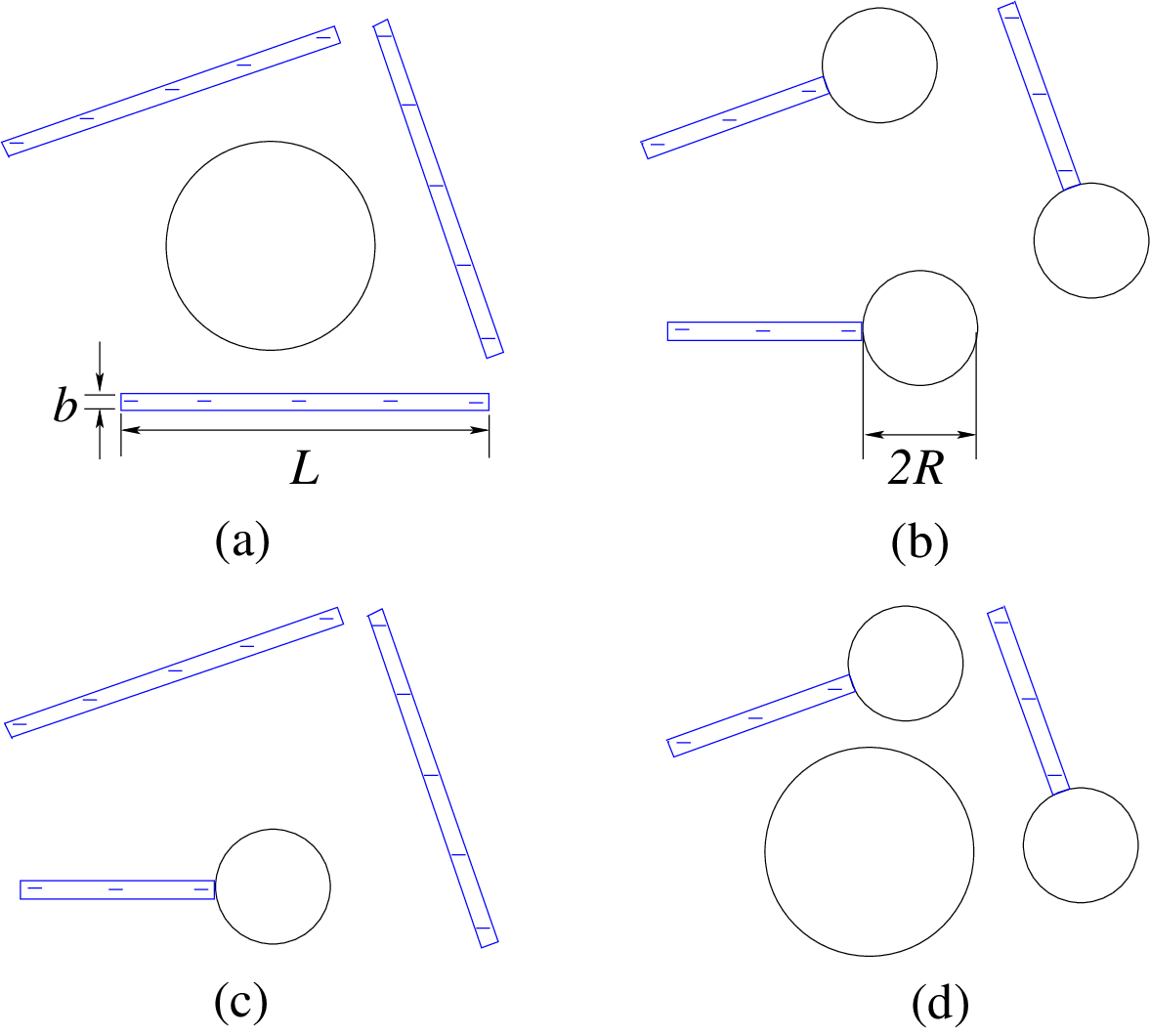}
\end{center}
\caption{Possible phases and phase coexistence due to
disproportionation. Symbols used here are explained in
Fig.~\ref{fig:PE}. (a) Coexistence of free complexes and a single
drop (inter-complex disproportionation): PCs disproportionate
themselves among PAs so that a portion of PA-PCs complexes are
neutral and condensed in a macroscopic drop, while the rest of
them are stronger charged and free. (b) The tadpole phase
(intra-complex disproportionation): PCs disproportionate
themselves within each PA-PCs complex to form a ``tadpole", with a
neutral condensed ``head" and a charged ``tail". (c) Coexistence
of tadpoles and free complexes. (d) Coexistence of tadpoles and a
single drop. In these two cases, inter and intra
disproportionation  are combined to achieve a lower free
energy.}\label{fig:condensate}
\end{figure}

In vicinity of $x=1$, the long-range Coulomb repulsion between
charged complexes competes with the short-range attraction due to
correlations. As $x$ increases, condensation starts at $x=x_1$. To
minimize the free energy, PCs can be redistributed among complexes
so that a portion of complexes are neutral and therefore condensed
in a macroscopic drop, while the rest of complexes become stronger
charged and stay free. This is called \emph{inter-complex
disproportionation}~\cite{Kabanov} or partial
condensation~\cite{Nguyen-reentrant} (see
Fig.~\ref{fig:condensate}a). It is essentially a coexistence of
the two phases: the single drop phase and the free complexes
phase. 

On the other hand, PCs can also disproportionate themselves
within each complex, which we call \emph{intra-complex
disproportionation} (Fig.~\ref{fig:condensate}b). PCs move closer
to one end of PA molecule, making one part of a PA molecule
neutral and condensed in a droplet, while the other part is even
stronger charged and not condensed. Unlike inter-complex
disproportionation,  this gives a new ``tadpole" phase, as far as
$L$ is finite, where $L$ is the length of a free PA-PCs complex
(see Fig.~\ref{fig:phase}). It is possible to combine the two ways
of disproportionation to accomplish even lower free energies (see
Fig.~\ref{fig:condensate}c,d). As a result, at
$r_s\rightarrow\infty$, we have the sequence of these phases as
shown on the horizontal axis of Fig.~\ref{fig:phase}.

\begin{figure}[ht]
\begin{center}
\includegraphics[width=0.8\textwidth]{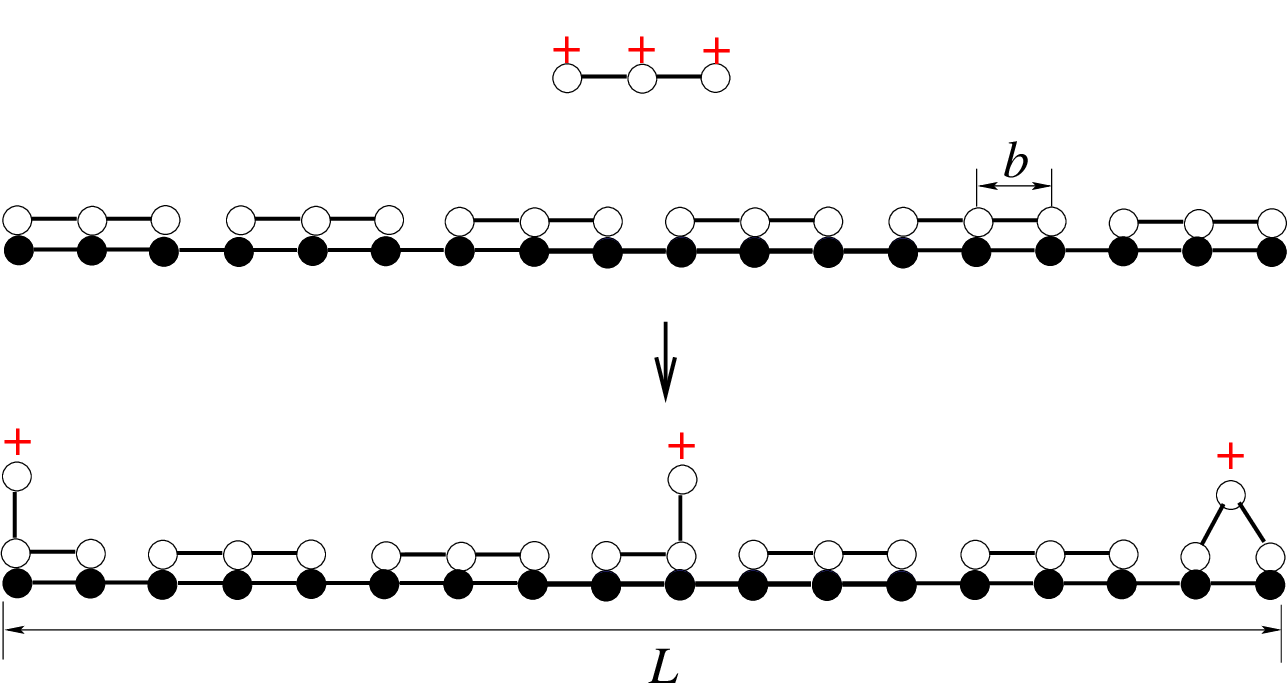}
\end{center}
\caption{An illustration of charge inversion of a PA molecule by
flexible PCs when they are both strongly charged. Negative PA
charges are shown by black dots. Positive PC charges are shown by
white dots. When a new PC molecule is adsorbed to a neutral PA-PCs
complex, its charge is fractionalized in mono-charges and its
Coulomb self-energy is eliminated by redistribution of all PCs in
the complex. In reality, the numbers of charges of PA and PC can
be much larger than numbers shown here. Imagine for example that
PA and PC have charges $-1000e$ and $+100e$.}\label{fig:fraction}
\end{figure}

What happens at $x>1$ side is almost the same, except that here
the number of PC is larger than necessary to neutralize all PAs
and each complex is strongly positively charged (charge inverted)
(Fig.~\ref{fig:PE}c). Let us briefly remind the nontrivial
mechanism of charge inversion at $x>1$
side~\cite{Nguyen-fraction}. We illustrate it in
Fig.~\ref{fig:fraction} for the model of strongly charged flexible
PA and PC, in which the distance between charges, $b$, is the same
for PA and PC molecules and is of the order of Bjerrum length
$l_B=7 \AA$ ($e^2/Dl_B=k_BT$, $D=80$ is the dielectric constant of
water). When a new PC molecule arrives at a neutral PA-PCs
complex, all PCs in the complex can rearrange themselves so that
the charge of this excessive PC is smeared in the whole complex
and the Coulomb self-energy of the PC is effectively reduced to
zero (Fig.~\ref{fig:fraction}). This elimination of the Coulomb
self-energy is essentially due to correlation of PCs in the
complex and can not be described by Poisson-Boltzmann mean field
approximation. We define $\mu_c<0$ ($c$ stands for ``correlation")
as the chemical potential related to the elimination of the
Coulomb self-energy of PC in the complex. Related to the charge of
PC, it acts as an external voltage overcharging PA. With
increasing $x$, the inverted charge of the complex increases. At
certain critical $x=x_{m}$ ($m$ stands for ``maximum charge
inversion") (see Fig.~\ref{fig:phase}), the maximum charge
inversion is achieved where $\mu_c$ is balanced by the Coulomb
repulsive energy of the complex to a PC. The topological structure
of the phase diagram is almost symmetric about $x=1$. The only
asymmetry appears at $x>x_m$. Here additional PCs are not
attracted to maximum charge-inverted PAs and stay free in the
solution (see Fig.~\ref{fig:phase}).

Now let us discuss screening by a monovalent salt. This screening
effectively cuts off the range of the Coulomb interaction at the
distance $r_s$. As $r_s$ decreases, first the long-range Coulomb
repulsion is reduced, while the short-range correlation induced
attraction is not affected. Accordingly, all the ranges of
condensation in the phase diagram become wider
(Fig.~\ref{fig:phase}). Eventually, in the limit of very small
$r_s$, the short-range correlational attraction is also screened
out and the macroscopic drop completely dissolves (such small $r_s$ are not shown in
Fig.~\ref{fig:phase}). In the intermediate range of $r_s$ we are
interested in, there are two major effects of screening. First,
the tadpole configuration disappears at certain very large $r_s = r_c$ given by Eq. (37)
(Fig.~\ref{fig:phase}). We will show that practically always $r_s < r_c$ 
or $b/r_s > b/r_c$ so that tadpoles are hardy observable.

Second effect of screening is that the single drop phase occupies
a finite range of $x$ around $x=1$. Namely, at a finite $r_s$,
as we show below in Eq. (41) this range widens with decreasing $r_s$ 
as $1/r_s^2$ (Fig.~\ref{fig:phase}).
Recall that at $r_s\rightarrow\infty$, the macroscopic drop should be 
neutral and therefore it exists only at $x=1$. If each condensed complex were
charged ($x\neq 1$), the total charge of the macroscopic drop
would be proportional to its volume and the Coulomb energy per
unit volume would be huge (proportional to its surface). On the
other hand, at $r_s$ much smaller than the size of the macroscopic
drop, the Coulomb energy per unit volume is constant. 
Therefore the macroscopic drop can tolerate some charge density 
and the range of the single drop phase in the phase diagram widens.

Currently our theory can be compared with experiments only
qualitatively since in many cases it is not clear, whether the
equilibrium state of the system is reached in experimental times
due to the slow kinetics. Also the interesting tadpole phase is
very hard to realize due to a very large critical value $r_c$.
In solutions of DNA with PC, charge
inversion of complexes is observed at $x > 1$~\cite{Kabanov,Budker}.
The size of condensed particles reaches maximum close to $x=1$
corresponding to the single drop phase in our phase diagram. When
$x\neq 1$, the size of condensed particles decreases in agreement
with our equilibrium phase diagram~\cite{Kabanov,Budker,van
Zanten}. In solutions of DNA with basic polypeptides, at $x<1$, it
is observed that DNA molecules exist simultaneously in two
distinct conformations, i.e., elongated conformation and condensed
conformation~\cite{Minagawa}. This corresponds to a phase
coexistence of free complexes and a single drop in our phase
diagram ($x_1<x<1$ in Fig.~\ref{fig:phase}). On the other hand,
the enhancement of condensation with the help of simple salt is
observed in Ref.~\cite{Budker}. Certain tadpole-like phases have
also been observed in experiments~\cite{Budker,Dunlap,Zinchenko},
although we doubt that they are equilibrium tadpole phases
discussed in this paper.

This paper is organized as follows. In Sec.~\ref{sec:nosalt}, we
discuss all possible phases when $r_s\rightarrow\infty$. We then
consider the role of screening by monovalent salt and get the
complete phase diagram in Sec.~\ref{sec:screen}. In
Sec.~\ref{sec:inverse}, we discuss $x>1$ side more carefully and
reveal another possibility for the phase diagram. In
Sec.~\ref{sec:PE}, we estimate parameters $\varepsilon$ and
$\mu_c$ microscopically in the case of strongly charged PA and PC.
In Sec.~\ref{sec:entropy}, we discuss the role of translational
entropy of PC in connection with previous
works~\cite{Nguyen-reentrant,Rui}. We conclude in
Sec.~\ref{sec:conclusion}.

\section{Phases in the absence of a salt}\label{sec:nosalt}
In this section we discuss all possible phases in the absence of
salt, i.e., $r_s\rightarrow\infty$. Although this
situation is not realistic, it serves as a starting point for more
complicated theory at finite $r_s$ (see Sec.~\ref{sec:screen}).
Here and below we focus on the role of the Coulomb interaction
and neglect all other interactions such as the hydrophobic one. We
focus on $x<1$ side and postpone the discussion of $x>1$ side till
Sec.~\ref{sec:inverse}.

There are three main phases in our system: the phase of a single drop
(Fig.~\ref{fig:PE}d), the phase of free complexes
(Fig.~\ref{fig:PE}b), and the tadpole phase
(Fig.~\ref{fig:phase}e). At $x=1$, there is no Coulomb repulsion 
between complexes, the
phase of a single drop is preferred due to the short range
correlation energy gain (see Fig.~\ref{fig:correlation}). At $x\ll
1$, the phase of free complexes is preferred because of strong
Coulomb repulsion between complexes. Then it is natural to
conclude that the tadpole phase, which is kind of combination of
the other two phases, appears in vicinity of $x=1$. To verify
this idea, we focus on possible phase coexistent regions. 
The phase of free complexes can coexist with the tadpole
phase, and the tadpole phase can coexist with the phase of a
single drop. By minimizing the free energies in these two regions
respectively, we can find the ranges of $x$ in which one or the
other phase is preferred, or the two phases coexist. This leads to
the phase diagram at $r_s\rightarrow\infty$.

\subsection{The free energies of the three phases \label{sub:each}}
For pedagogical reason, let us start from the free energy of the
three phases. We first notice that the Coulomb energy of each
charged complex can be calculated by considering it as a conductor
with certain capacitance. To see this, let us recall that the
total chemical potential of a PC molecule adsorbed in PA is given
by
\begin{equation}
\mu=\mu_c+q\phi.
\end{equation}
Here the first term is the correlational chemical potential, and
the second term is the electric energy of a PC given by the local
electric potential $\phi$ in the complex. Since both $\mu$ and
$\mu_c$ are the same along the complex, $\phi$ must be the same in
the complex. It is in this sense that the PA-PCs complex can be
considered as a conductor and the concept of capacitance can be
used to calculate its Coulomb energy.

We denote $N$ as the total number of PA molecules in the solution.
For the phase of free complexes (see Fig.~\ref{fig:PE}b), we have
\begin{equation}
F=N\left[\frac{(nq-Q)^2}{2C}+nE(n)\right].
\end{equation}
Here $C$ is the capacitance of a free complex, $-Q$ and $q$ are
the bare charges of PA and PC, $n$ is the number of PC in each
free PA-PCs complex, and $E(n)<0$ is the correlation energy of a
PC in a complex as a function of $n$. In this expression, the
first term is the Coulomb self-energy of free complexes. The
second term is the negative correlation energy of PCs in free
complexes. We emphasize again that we study very long and strongly
charged PC and PA such that their translational entropies are
negligible. Since all PCs are adsorbed to PAs, the net charge of
each free complex, $(nq-Q)$, is equal to $(x-1)Q$. We will show in
Sec.~\ref{sec:PE} that near the phase coexistence region,
$|n-n_i|\ll n_i$, where $n_i=Q/q$ is the number of PC in a neutral
complex. Consequently, $\mu_c(n)=\partial [nE(n)]/\partial n$ is
approximately equal to its value $\mu_c$ at $n=n_i$. Furthermore,
it is convenient to consider the average free energy of each
complex, $f=F/N$, instead of $F$. Finally, we rewrite the free
energy as
\begin{equation}
f=\frac{(x-1)^2Q^2}{2C}+\frac{(x-1)Q}{q}\mu_c+n_iE(n_i),\label{eq:fcomplex}
\end{equation}
where $C$ is given by
\begin{equation}
C=\frac{DL}{2\ln(L/b)}\label{eq:C}
\end{equation}
as the capacitance of a cylindrical free PA-PC complex capacitor of the length $L$ and  $D=80$ is the dielectric
constant of water.

Using similar notations, for a phase of a single drop (see
Fig.~\ref{fig:PE}d), we have
\begin{equation}
f=n_iE(n_i)+\varepsilon. \label{eq:fdrop}
\end{equation}
Recall that each complex in the macroscopic drop is neutral. The
expression includes the negative correlation energy of PCs in
neutral complexes, and the negative correlation energy of a
neutral complex due to condensation, $\varepsilon$.

For a tadpole phase (see Fig.~\ref{fig:PE}e), we denote $z$ as the
fraction of the tail part of each complex. We have
\begin{equation}
f(z)=\frac{(x-1)^2Q^2}{2C_t}
+(1-z)\varepsilon\left(1-\frac{3b}{R}\right)+\frac{(x-1)Q}{q}\mu_c
+n_iE(n_i).\label{eq:ftadpole}
\end{equation}
This free energy is almost a combination of the last two free
energies except two differences. First, the capacitance of a
tadpole is determined by its tail:
\begin{equation}
C_t=\frac{DzL}{2\ln(zL/b)}. \label{eq:Ct}
\end{equation}
Second, the correlation energy due to condensation is gained only
in the ``head" of the tadpole. Therefore there is an additional
factor $(1-z)$ in the second term. The surface energy of the head is
also included in the second term~\cite{ab}. $R$ is the radius of
the head of the tadpole (see Fig.~\ref{fig:condensate}b), given by
\begin{equation}
R=\left[\frac{3}{16}(1-z)b^2L\right]^{1/3}.\label{eq:R}
\end{equation}

\subsection{The phase diagram at $r_s\rightarrow\infty$\label{sub:diagram}}
Now we are ready to discuss the free energy in the mixed phases and get the phase diagram at
$r_s\rightarrow\infty$. We first consider the coexistence of the
phase of free complexes and the tadpole phase (see
Fig.~\ref{fig:condensate}c). We denote $y$ as the fraction of
tadpoles in the mixture. Combining Eqs.~(\ref{eq:fcomplex})
and~(\ref{eq:ftadpole}), we have
\begin{equation}
f(y,z)=\frac{(x-1)^2Q^2}{2[yC_t+(1-y)C]}
+y(1-z)\varepsilon\left(1-\frac{3b}{R}\right)+\frac{(x-1)Q}{q}\mu_c
+n_iE(n_i).\label{eq:fyz}
\end{equation}
Here the first term is the Coulomb energy of the system of free
complexes and tadpoles with capacitence $C$ and $C_t$
correspondingly. These capacitances are additive because all
complexes are in equilibrium with respect of exchange of PC and
the electric potential of all complexes is the same (see
discussion at the beginning of the subsection ~\ref{sub:each}).

To minimize this free energy with respect of two variables $y$ and
$z$, it is convenient to separate the free energies into two parts
\begin{equation}
f(y,z)/|\varepsilon|=\left[\alpha\frac{(1-x)^2\ln
(L/b)}{1-y+yz}-y(1-z)\right]+\left[\alpha\frac{(1-x)^2yz\ln
z}{(1-y+yz)^2}+\frac{5y(1-z)^{2/3}}{(L/b)^{1/3}}\right],\label{eq:fyz1}
\end{equation}
where
\begin{equation}
\alpha=\frac{Q^2}{DL|\varepsilon|}\label{eq:alpha}
\end{equation}
is in order of $1$ for strongly charged PA and PC (see
Sec.~\ref{sec:PE}), and Eqs.~(\ref{eq:C}),~(\ref{eq:Ct})
and~(\ref{eq:R}) have been used. In this expression, the last two
terms of Eq.~(\ref{eq:fyz}) have been neglected since they are
independent of $y,z$. The first square bracket contains the main
terms of the Coulomb energy and the correlation energy, in which
$\ln zL$ in $C_t$ has been replaced by $\ln L$ and the surface
energy has been ignored. The second square bracket contains
correction terms in the first order to make up the two neglects in
the main terms. We will see below that the small parameter we used
to expand this free energy is $\ln(L/b)/(L/b)^{1/3}$, for $L\gg
b$.

We first minimize the main terms in Eq.~(\ref{eq:fyz1}) and get
\begin{equation}
y(1-z)=1-(1-x)\sqrt{\alpha\ln (L/b)}.
\end{equation}
Putting it back to the correction terms in Eq.~(\ref{eq:fyz1}) and
taking the minimum, we get
\begin{equation}
z=1-\left[\frac{5\ln(L/b)}{3(L/b)^{1/3}}\right]^{3/4}.
\end{equation}
Combining the above two conditions, we have
\begin{equation}
y=\left[1-(1-x)\sqrt{\alpha\ln(L/b)}\right]\left[\frac{3(L/b)^{1/3}}{5\ln
(L/b)}\right]^{3/4}.
\end{equation}
Specifically, considering $y=0$ and $y=1$, we get the boundaries
of the phase coexistent region (see Fig.~\ref{fig:phase})
\begin{eqnarray}
1-x_1(\infty)&=&\frac{1}{\sqrt{\alpha\ln(L/b)}},\label{eq:x10}\\
1-x_2(\infty)&=&\frac{1}{\sqrt{\alpha\ln(L/b)}}
\left\{1-\left[\frac{5\ln(L/b)}{3(L/b)^{1/3}}\right]^{3/4}\right\}.\label{eq:x20}
\end{eqnarray}
Here and below $x(\infty)$ means $x(r_s\rightarrow\infty)$. When
$x$ increases from $x_1(\infty)$ to $x_2(\infty)$, $y$ increases
from $0$ to $1$ linearly. This is actually the level
rule~\cite{Landau} and the sign of a first order phase transition.

For a very long PA, $L\gg b$, parameter $\ln(L/b)/(L/b)^{1/3}\ll
1$. Therefore $z$ is close to 1, the free energy expansion in
Eq(\ref{eq:fyz1}) is self-consistent. Also $x_1(\infty)$ is close
to 1 and $x_2(\infty)$ is close to $x_1(\infty)$.

We consider the other mixed phases (see
Fig.~\ref{fig:condensate}d) following the same approach. Combining
Eqs.~(\ref{eq:fdrop}) and~(\ref{eq:ftadpole}), we have
\begin{equation}
f(y,z)=\frac{(x-1)^2Q^2}{2yC_t}
+y(1-z)\varepsilon\left(1-\frac{3b}{R}\right)+(1-y)\varepsilon
+\frac{(x-1)Q}{q}\mu_c +n_iE(n_i).\label{eq:fyz2}
\end{equation}
The free energy similar to Eq.~(\ref{eq:fyz1}) is
\begin{equation}
f(y,z)/|\varepsilon|=\left[\alpha\frac{(1-x)^2\ln
(L/b)}{yz}+yz-1\right]+\left[\alpha\frac{(1-x)^2\ln
z}{yz}+\frac{5y(1-z)^{2/3}}{(L/b)^{1/3}}\right].\label{eq:fyz3}
\end{equation}
Minimizing the main terms (in the first square bracket), we get
\begin{equation}
yz=(1-x)\sqrt{\alpha\ln(L/b)}.\label{eq:y}
\end{equation}
Putting it back to the correction terms (in the second square
bracket) and taking the minimum, we get
\begin{equation}
z=\frac{5\ln(L/b)}{(L/b)^{1/3}}.
\end{equation}
Combining the two conditions, we have
\begin{equation}
y=\frac{(1-x)\sqrt{\alpha}(L/b)^{1/3}}{5\sqrt{\ln(L/b)}}.
\end{equation}
Therefore, the boundaries of this mixture are given by $y=1$ and  $y=0$ (see
Fig.~\ref{fig:phase})
\begin{eqnarray}
1-x_3(\infty)&=&\frac{1}{\sqrt{\alpha\ln(L/b)}}\frac{5\ln(L/b)}{(L/b)^{1/3}},\\
1-x_4(\infty)&=&0.
\end{eqnarray}
We see that the same small parameter $\ln(L/b)/(L/b)^{1/3}$ works
here. Again, the phase transition from the tadpole phase to the
phase of a single drop is the first order one.

\subsection{Why not other phases}
In this subsection, we argue that several other possible phases or
phase coexistence do not appear in the phase diagram at
$r_s\rightarrow\infty$.

One possibility would be that the tadpole phase does not appear at
all, but the coexistence of the other two phases shows up in
vicinity of $x=1$.  This is called inter-complex
disproportionation (Fig.~\ref{fig:condensate}a). We showed in
Ref.~\cite{Rui3} that this coexistence has a higher free energy
than the tadpole phase in some range of $x<1$. As far as the
tadpole phase exist, it immediately follows that the only phase
coexistent regions can exist are those involving tadpoles (see
Fig.~\ref{fig:condensate}c,d). And the inter-complex
disproportionation region (Fig.~\ref{fig:condensate}a) has nowhere
to show up.

\begin{figure}[ht]
\begin{center}
\includegraphics[width=0.4\textwidth]{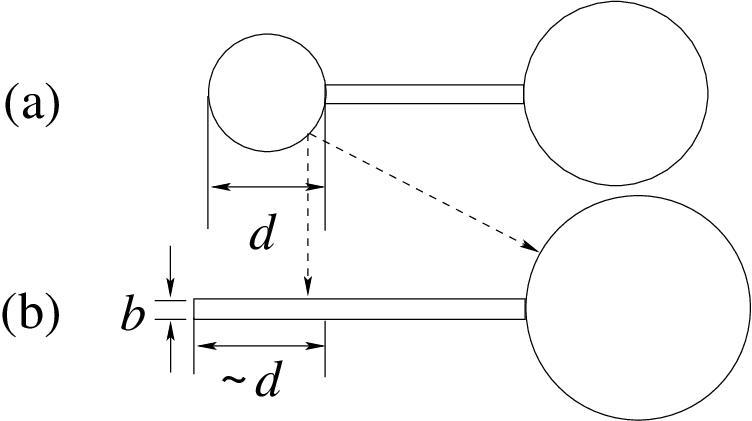}
\end{center}
\caption{Comparison of two-heads and one-head configurations. (a):
a two-head configuration. The diameter of the smaller head is $d$.
(b): a one-head configuration made from (a) by combining the two
heads then releasing tail of length order of $d$ from the head,
such that the two configurations have same capacitance. The
Coulomb energy is the same for the two configurations, but the
surface energy is higher in (a).}\label{fig:heads}
\end{figure}

The second possibility is a phase in which every complex has more
than one ``head" and therefore forms a necklace structure. Let us
argue that the one-head-one-tail tadpole configuration is the best
for intra-complex disproportionation. To do this, we have to
include the capacitances of heads which gives a small correction
to Eq.~(\ref{eq:Ct}). First of all, let us show that for a complex
with given charge, one head is better than two heads. Consider an
arbitrary two-heads configuration with the diameter of the smaller
head $d$ (Fig.\ref{fig:heads}a). We can always construct an
one-head configuration from it by combining the two heads then
releasing additional tail of the order of $d$ from the head in
such a way that the capacitances of the two configurations are the
same (Fig.\ref{fig:heads}b). The total free energy consists of the
long-range Coulomb energy and the short-range correlation energy
in droplets. For the two configurations, the Coulomb energies are
the same since the capacitances are equal. But the surface energy
is higher in the two-heads configuration since the surface is much
larger. Thus, for any two-heads configuration, we can always find
a one-head configuration with lower energy. By similar argument,
obviously a configuration with many heads along the complex is
even worse. Furthermore, the single head always prefers to be at
the end of the tail. This can be understood by considering a
metallic stick with fixed charge on it. The electric field is
larger at the end of the stick than in the middle. Therefore to
reduce the Coulomb energy, it is better to put a metallic sphere
at the end of the stick to make field there smaller.

\begin{figure}[ht]
\begin{center}
\includegraphics[width=0.8\textwidth]{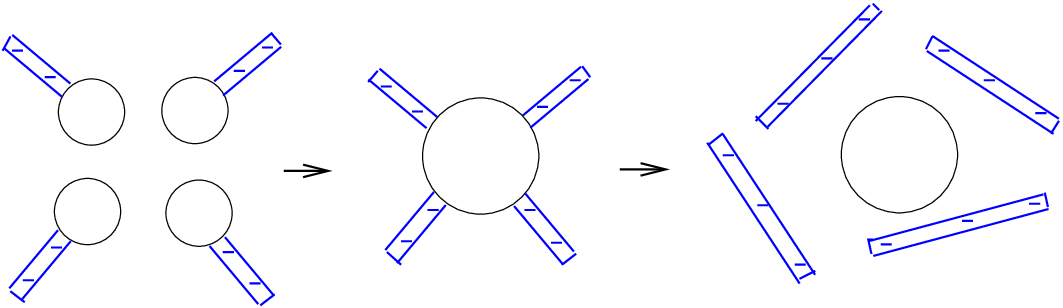}
\end{center}
\caption{An illustration of instability of hypothetic micelle-like
droplets in a PA-PC system. Symbols are explained in
Fig.~\ref{fig:PE}. If tadpoles are not stable, one could think
that they merge into micelles. But micelles would immediately
break down to reduce Coulomb energy.}\label{fig:merge}
\end{figure}
The third possibility is a phase in which every few complexes come
together to form a micelle-like object (see Fig.~\ref{fig:merge}).
However, micelle-like droplets are not stable. For example,
suppose several charged PA-PCs tadpoles would gain energy merging
into a micelle. Instead of sharing one head by several tails, to
reduce the Coulomb energy, it is better to break the micelle into
one tadpole and several tails. Repeating this process, certainly
one ends up with a two phase coexistence (see
Fig.~\ref{fig:merge}).

Finally, we want to point out that the necklace phase and the
micelle phase are always worse in free energies than the tadpole
phase. This is not only true at $r_s\rightarrow\infty$. One can
show that they do not exist in the case of finite $r_s$ too.

\section{Screening by monovalent salt.}
\label{sec:screen}
In this section, we consider a more realistic situation of finite
$r_s$ and get the complete phase diagram in a plane of $b/r_s$ and
$x$ (see Fig.~\ref{fig:phase}). Again we focus on $x<1$ side. The
$r_s$ range we are interested in is $b\ll r_s\ll L$. As $r_s$
becomes smaller than $L$, the long-range Coulomb repulsion is
first reduced while the short-range correlational attraction is
not affected. Eventually, when $r_s<b$, the short-range
correlational attraction is also screened out and the macroscopic
drop completely dissolves but we do not consider such salt
concentrations. In the range of our interest, $b\ll r_s\ll L$,
there are two major implications of screening (see
Fig.~\ref{fig:phase}). First, the phase of single drop grows up.
Second, the tadpole phase and related phase coexistent regions are
destroyed.

\begin{figure}[ht]
\begin{center}
\includegraphics[width=0.5\textwidth]{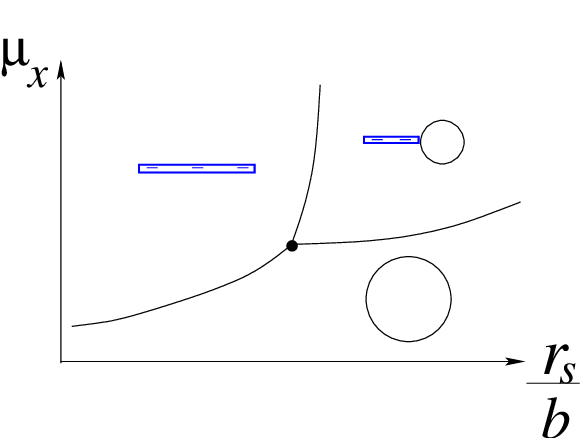}
\end{center}
\caption{Schematic phase diagram on a plane of $\mu_x$ and $r_s$.
Here $\mu_x$ is a chemical potential which is conjugate to
$x$.}\label{fig:triple}
\end{figure}

Let us first discuss the general confinement to the phase diagram
given by the Gibbs' phase rule~\cite{Landau}. Suppose the number
of coexistent phases is $m$. Then for our two substances system,
the number of independent equations following the condition of
equal chemical potentials is $2(m-1)$. The number of unknowns is
$m+1$. For example, they can be chosen as the charge ratio $x$ in
each phase and the common parameter $b/r_s$ shared by all
coexistent phases. To have a solution to the equations, we require
$2(m-1)\leq m+1$. This gives $m\leq 3$. That is to say, the number
of coexistent phases cannot be more than 3. It is convenient to
define a chemical potential
\begin{equation}
\mu_x=\frac{\partial f}{\partial x},
\end{equation}
and consider the phase diagram on a plane of $\mu_x$ and $b/r_s$
(see Fig.~\ref{fig:triple}). When three phases coexist, all
variables are completely determined. It corresponds to a point in
the phase diagram (the triple point). When two phases coexist,
there is one thermodynamic degree of freedom. It corresponds to a
line in the phase diagram. If we change variable from $\mu_x$ to
its conjugate variable, $x$, we get our phase diagram
(Fig.~\ref{fig:phase}). One can easily show that in this phase
diagram, three phases coexist on a line, and two phases coexist in
a region.

The phase diagrams we get is very much similar to the phase
diagram of water~\cite{Landau}. Indeed, if we consider $\mu_x$ as
pressure, $r_s/b$ as temperature, and $x$ as volume, then
Fig.~\ref{fig:triple} is like the PT diagram, and
Fig.~\ref{fig:phase} is like the TV diagram of water. Notice that
in our system, real temperature or pressure plays no role. The
system is always at room temperature. And the pressure is
negligibly small, since in comparison of the Coulomb energy, we
have completely ignored the translational entropies of PA and PC,
which is related to the volume and the pressure of the system.

Having figured out the topological structure of the phase diagram,
let us determine all phase boundaries by considering the three
two-phase coexistent regions. We start from the mixture of
tadpoles and free complexes (Fig.~\ref{fig:condensate}c). In the
presence of monovalent salt, the free energy is still given by
Eq.~(\ref{eq:fyz}),  but the expressions of $C$ and $C_t$ can be
different. Let us first assume that $zL<r_s\ll L$. Then $C_t$ is
still given by Eq.~(\ref{eq:Ct}), while $C$ is now
\begin{equation}
C=\frac{DL}{2\ln(r_s/b)}.\label{eq:C2}
\end{equation}
Following the same procedure as in subsection~\ref{sub:diagram},
we get
\begin{eqnarray}
z&=&\frac{r_s}{2.7L}\exp\left[-\frac{5\ln(r_s/b)}{3(L/b)^{1/3}}\right],
\label{eq:z}\\
y&=&\frac{1-(1-x)\sqrt{\alpha\ln(r_s/b)}}{1-z}.
\end{eqnarray}
At $y=0$ and $y=1$, we get the boundaries of this coexistent
region (see Fig.~\ref{fig:phase})
\begin{eqnarray}
1-x_1(r_s)&=&\frac{1}{\sqrt{\alpha\ln(r_s/b)}}, \\
1-x_2(r_s)&=&\frac{r_s}{2.7L\sqrt{\alpha\ln(r_s/b)}}
\exp\left[-\frac{5\ln(r_s/b)}{3(L/b)^{1/3}}\right].
\end{eqnarray}
In the limiting case of $r_s\rightarrow L$, these equations go
back to Eqs.~(\ref{eq:x10}) and (\ref{eq:x20}).

If $r_s\ll zL$, not only $C$, but also $C_t$ gets a new expression
\begin{equation}
C_t=\frac{DzL}{2\ln(r_s/b)}.\label{eq:ct2}
\end{equation}
In this case, one can check that the free energy given by
Eq.~(\ref{eq:fyz}) does not have a minimal extremum. Indeed, the
Coulomb energy of the system becomes so short-ranged that there is
almost no difference to distribute all charges to free complexes
or tadpoles. The surface energy of heads dominates and the
coexistence of free complexes and the single drop is preferred
(see Fig.~\ref{fig:phase}).

We now consider the coexistence of tadpoles with the single drop
(Fig.~\ref{fig:condensate}d). In the case of finite $r_s$, the
free energy given by Eq.~(\ref{eq:fyz2}) should be revised to
\begin{equation}
f(y,z)=\frac{(x-1)^2Q^2}{2[yC_t+(1-y)C^{\prime}]}
+y(1-z)\varepsilon\left(1-\frac{3b}{R}\right)+\frac{(x-1)Q}{q}\mu_c
+n_iE(n_i).\label{eq:fyz4}
\end{equation}
Here $C^{\prime}$ represents the capacitance of a condensed
complex in the single drop. As discussed in
Sec.~\ref{sec:introduction}, when $r_s\rightarrow\infty$, the
macroscopic condensate is almost neutral and the phase of single
drop exist only at $x=1$. For finite $r_s$, the macroscopic drop
can tolerate certain charge density. Therefore $C^{\prime}$
appears. In order to calculate $C^{\prime}$, we assume that the
macroscopic drop is uniformly charged~\cite{Rui2}. If the charge
density of the macroscopic drop is $\rho$ and the charge of each
complex is $\pi b^2 L\rho/4$~\cite{ab}, the electrical potential
of the macroscopic drop is
\begin{equation}
\phi=\int_0^\infty\frac{\rho e^{-r/r_s}}{Dr}4\pi r^2dr=\frac{4\pi
r_s^2\rho}{D},
\end{equation}
This gives
\begin{equation}
C^{\prime}=\frac{\pi b^2
L\rho}{4\phi}=\frac{Db^2L}{16r_s^2}.\label{eq:C'}
\end{equation}
When $r_s\rightarrow\infty$, the capacitance
$C^{\prime}\rightarrow 0$ as expected.

In the case of $zL<r_s\ll L$, $C_t$ is still given by
Eq.~(\ref{eq:Ct}). Minimizing the free energy as in
subsection~\ref{sub:diagram}, we get
\begin{equation}
z=\frac{5\ln(L/b)}{(L/b)^{1/3}}+\frac{\ln^2(L/b)}{8(r_s/b)^2}.
\end{equation}
And $y$ is still given by Eq.~(\ref{eq:y}). Setting $y=0$ and
$y=1$, the boundaries of the coexistent region are (see
Fig.~\ref{fig:phase})
\begin{eqnarray}
1-x_3(r_s)&=&\frac{1}{\sqrt{\alpha\ln(L/b)}}
\frac{5\ln(L/b)}{(L/b)^{1/3}}\left[1+\frac{(L/b)^{1/3}\ln(L/b)}{40(r_s/b)^2}\right],\\
1-x_4(r_s)&=&0.
\end{eqnarray}

In the other case of $r_s\ll zL$, $C_t$ is given by
Eq.~(\ref{eq:ct2}). Again the free energy Eq.~(\ref{eq:fyz4}) has
minimum at $z=1$. This leads to the coexistence of free
complexes and the single drop is preferred (see
Fig.~\ref{fig:phase}).

We are now ready to determine the position of the line at which
three phases coexist in Fig.~\ref{fig:phase} (it is also the upper
boundary of the tadpole phase and the two mixed phases). 
The critical $r_s=r_c$, is determined by the 
point where of $x_2(r_s) = x_3(r_s)$. In the leading order, we have
\begin{equation}
r_c > L^{2/3}b^{1/3}\ln(L/b).
\end{equation}
Thus, $r_c$ is very large. For example, for $b=7$ $\AA$ and $L=1000b$, $r_c >
690b$. This means that the tadpole phase and all mixed phases related with it 
are not relevant to typical experiments. 

Finally we consider the coexistence of free complexes and the
single drop (see Fig.~\ref{fig:condensate}a) for realistic $r_s < r_c$.
Combining Eqs.~(\ref{eq:fcomplex}) and (\ref{eq:fdrop}), we have
\begin{equation}
f(y)=\frac{(x-1)^2Q^2}{2[yC+(1-y)C^{\prime}]} +(1-y)\varepsilon
+\frac{(x-1)Q}{q}\mu_c +n_iE(n_i),\label{eq:fyz5}
\end{equation}
where $C$ is given by Eq.~(\ref{eq:C2}). Minimizing the free
energy, we have
\begin{equation}
(x-1)^2\left[\frac{1}{\ln(r_s/b)}-\frac{b^2}{8r_s^2}\right]
=\frac{D|\varepsilon|
L}{Q^2}\left[\frac{y}{\ln(r_s/b)}+\frac{(1-y)b^2}{8r_s^2}\right]^2
\label{eq:eqcondsc}.
\end{equation}
For $y=1$ and $y=0$ we get
\begin{eqnarray}
1-x_1(r_s<r_c)&=&\frac{1}{\sqrt{\alpha\ln(r_s/b)}},\label{eq:x1} \\
1-x_4(r_s<r_c)&=&\frac{b^2}{8r_s^2}\sqrt{\frac{\ln(r_s/b)}{\alpha}}.
\end{eqnarray}
Here we put $r_s<r_c$ as the argument to remind us that these
expressions are meaningful only at $r_s<r_c$. We see that the
width of the single drop phase grows proportionally to $1/r_s^2$
with decreasing $r_s$ (see Fig.~\ref{fig:phase}). When
$r_s\rightarrow r_c$, $x_4(r_s)\rightarrow 1$ as expected.

\section{Possible new phases at $x>1$ side}\label{sec:inverse}
In this section, we discuss the phase diagram at $x>1$ side. At
$x>1$, the number of PC molecules is more than enough to
neutralize all PA molecules. The signs of charges of free
complexes or the tails of tadpoles are inverted to positive. In
spite of this difference, all physics we discussed in the last two
sections remain valid. Therefore one expects that the topological
structure of the phase diagram is symmetric about $x=1$. And the
boundaries $x_i^{\prime}(r_s)$ (see Fig.~\ref{fig:phase}) satisfy
\begin{equation}
x^{\prime}_i-1=1-x_i,
\end{equation}
where $i=1,2,3,4$.

However, a little asymmetry exists since the charge inversion
process cannot go forever, but reaches its maximum value at
certain critical value of $x$, $x_m$. $x_m$ is determined from the
balance of the gain in the correlation energy with the overall
Coulomb repulsive energy for a PC molecule. For example, for a
free complex, we have
\begin{equation}
|\mu_c|=\frac{(x_m-1)Qq}{C},\label{eq:mu}
\end{equation}
where each free complex carries net charge $(x_m-1)Q$. In this
equation, the left hand side is the magnitude of the gain in
correlation energy, and the right hand side is the Coulomb
repulsive energy given by the net charge of the complex. At
$x>x_m$, extra PC molecules are not attracted to the complexes,
but stay free in solution (see Fig.~\ref{fig:phase}). Using
Eq.~(\ref{eq:C2}), we get
\begin{equation}
x_m(r_s)=1+\frac{DL|\mu_c|}{2Qq\ln(r_s/b)}.\label{eq:xo}
\end{equation}
When $r_s\rightarrow\infty$,
\begin{equation}
x_m(r_s)=1+\frac{DL|\mu_c|}{2Qq\ln(L/b)}.\label{eq:xoo}
\end{equation}

Furthermore, the existence of $x_m$ gives another possibility to
the phase diagram. For our purpose, it is enough to consider the
simple case where the free complexes and the single drop coexist
(see Fig.~\ref{fig:condensate}a). One can show that the conclusion
is the same for phase coexistence involving tadpoles.

\begin{figure}[ht]
\begin{center}
\includegraphics[width=1\textwidth]{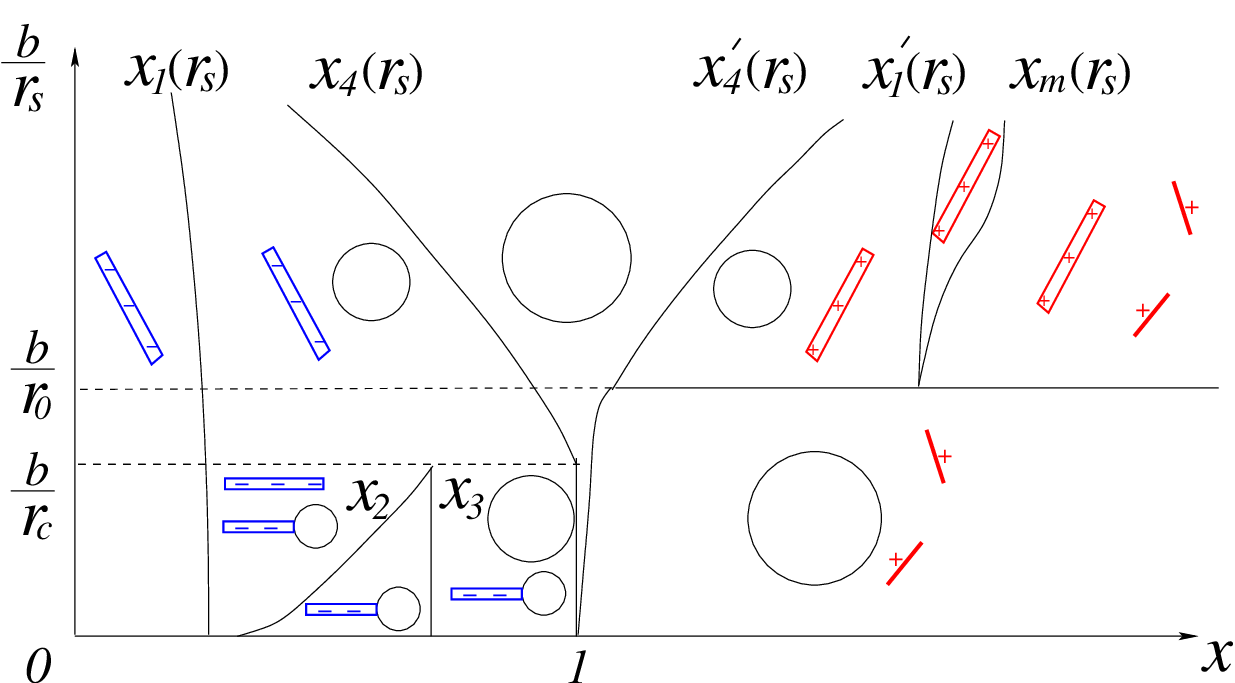}
\end{center}
\caption{Phase diagram of solution of PA and PC in the case when
at $r_s\rightarrow\infty$, we formally have $x_1^{\prime}>x_m$.
The meaning of axes and symbols are the same as
Fig.~\ref{fig:phase}. }\label{fig:phasehybrid}
\end{figure}

Indeed, what we discussed above is self-consistent if
$x^{\prime}_1<x_m$. But what if $x^{\prime}_1>x_m$? To answer this
question, we first notice that $x_1^{\prime}$ has another physical
meaning. In Eq.~(\ref{eq:eqcondsc}), for the accuracy we needed,
keeping the first term on each side, we get the net charge of each
free complex in the phase coexistent region
\begin{equation}
\frac{(x-1)Q}{y}=\frac{Q}{\sqrt{\alpha\ln(r_s/b)}}.
\end{equation}
According to Eq.~(\ref{eq:x1}), this net charge is equal to
$(x^{\prime}_1-1)Q$. When $x^{\prime}_1>x_m$, to keep the phase
coexistence, each free complex should carry charge
$(x_1^{\prime}-1)Q$ while it is only allowed to carry $(x_m-1)Q$
because of the finite correlation chemical potential $\mu_c$. In
this situation, $x^{\prime}_1$ loses its physical meaning and the
charge of each free complex saturates at the maximum value
$(x_m-1)Q$. The free energy of Eq.~(\ref{eq:fyz5}) should be
revised to
\begin{equation}
f(y)=y\frac{(x_m-1)^2Q^2}{2C}+y\frac{(x_m-1)Q}{q}\mu_c
+(1-y)\varepsilon+n_iE(n_i). \label{eq:fy}
\end{equation}
Here for the purpose of discussion, the second order term related
to $C^{\prime}$ has been ignored. With help of Eqs.~(\ref{eq:x1})
and~(\ref{eq:mu}), this free energy can be written as
\begin{equation}
f(y)=-y\frac{(x_m-1)^2Q^2}{2C}-(1-y)\frac{(x_1^{\prime}-1)^2Q^2}{2C}
+Nn_iE(n_i). \label{eq:free1}
\end{equation}
Clearly, when $x^{\prime}_1>x_m$, the minimum of $f(y)$ is reached
at $y=0$. Therefore at $x>1$ side, we arrive at a phase of total
condensation in which all complexes are condensed but some PCs are
free. This leads to a different phase diagram
(Fig.~\ref{fig:phasehybrid}).

It is convenient to classify all PA-PC systems into two
categories. In the first category,
$x_1^{\prime}(\infty)<x_m(\infty)$, and we have phase diagram
Fig.~\ref{fig:phase}. In the second category,
$x_1^{\prime}(\infty)>x_m(\infty)$, and we have phase diagram
Fig.~\ref{fig:phasehybrid}. Interestingly, both two categories are
realistic as we will see in Sec.~\ref{sec:PE}. Notice that in the
second case, decreasing $r_s$ to a critical value, $r_0$,
eventually leads to inversion of inequality $x_1^{\prime}>x_m$
(see Fig.~\ref{fig:phasehybrid}). We discuss this effect in detail
in the next section after we estimate $\varepsilon$ and $\mu_c$
microscopically.

In the case when $x_1^{\prime}(\infty)>x_m(\infty)$, the phase of
single drop expands around $x=1$ with growing $L/r_s$ at $x>1$
side too (see Fig.~\ref{fig:phasehybrid}). We calculate the
boundary of this phase with the phase of total condensation at
$r_s>r_0$. What we need to find out is just how many excessive PCs
the macroscopic drop can tolerate at finite $r_s$. Applying the
condition of maximum charge inversion to a condensed complex,
similarly to Eq.~(\ref{eq:mu}), we have
\begin{equation}
|\mu_c|=\frac{(x^{\prime}_s-1)Qq}{C^{\prime}}.
\end{equation}
Therefore
\begin{equation}
x^{\prime}_1(r_s<r_0)=1+\frac{D|\mu_c|b^2L}{16Qqr_s^2}.\label{eq:xs3}
\end{equation}
At $r_s<r_0$, we arrive at inequality $x_1^{\prime}(r_s)<x_m(r_s)$
and Eq.~(\ref{eq:x1}) gives the boundary of the phase of single
drop (see Fig.~\ref{fig:phasehybrid}). We will discuss the
transition at $r_s=r_0$ in detail in the next section.

\section{Phase diagram of Strongly charged polyelectrolytes}\label{sec:PE}
In this section, we consider a simple system where linear charge
densities of PA's and PC's are equal. Both of them are strongly
charged such that every monomer carries a fundamental charge $e$
and $e^2/Db\simeq k_BT$ (see Fig.~\ref{fig:fraction}). We estimate
parameters $\varepsilon$ and $\mu_c$ microscopically and choose
from the two phase diagrams shown in Figs.~\ref{fig:phase}
and~\ref{fig:phasehybrid}.

We first consider the case of $r_s\rightarrow\infty$. As discussed
in Sec.~\ref{sec:introduction}, $\mu_c$ is equal to the Coulomb
self-energy of a PC,
\begin{equation}
\mu_c=-\frac{qe}{Db}\ln(q/e).\label{eq:muc}
\end{equation}
On the other hand, when neutral PA-PCs complexes condense, they
form a strongly correlated liquid.  Monomers of two PEs locally
form NaCl-like structure such that the energy in order of
$-e^2/Db$ per monomer is gained
(Fig.~\ref{fig:correlation})~\cite{ab}. Therefore
\begin{equation}
\varepsilon\simeq -\frac{Qe}{Db}.
\end{equation}
We see that indeed $\alpha$ defined by Eq.~(\ref{eq:alpha}) is in
order of 1.

Substituting $L=Qb/e$ into Eqs.~(\ref{eq:x10}) and
~(\ref{eq:xoo}), we have
\begin{eqnarray}
x_1^{\prime}(\infty)&\simeq&1+\frac{1}{\sqrt{\ln(Q/e)}},\\
x_m(\infty)&=&1+\frac{\ln(q/e)}{2\ln(Q/e)}.\label{eq:xov}
\end{eqnarray}
Accordingly, we get the critical value of $q$ at which
$x_1^{\prime}=x_m$,
\begin{equation}
q_c=\exp\left(2\sqrt{\ln(Q/e)}\right).
\end{equation}
When $q>q_c$, or PC is long enough, $x_1^{\prime}<x_m$, we have
the phase diagram of reentrant condensation (horizontal axis of
Fig.~\ref{fig:phase}). When $q<q_c$, or PC is relatively short,
$x_1^{\prime}>x_m$, we have the asymmetric phase diagram with
total condensation at $x>1$ side (horizontal axis of
Fig.~\ref{fig:phasehybrid}). The possibility of having two
different phase diagrams for different $q$ is related to the
interplay of the logarithms in Eqs.~(\ref{eq:muc})
and~(\ref{eq:xov}). For fixed $Q$, when $q$ increases, $x_m$
increases but $x_1^{\prime}$ is fixed. Therefore we can have
either $x_m<x_1^{\prime}$ or $x_m>x_1^{\prime}$ by changing $q$.
Notice that $q_c$ is exponentially smaller than $Q$.

Now let us consider the effect of screening by monovalent salt. We
are interested in the case of $r_s\gg b$ when the short-range
correlation is not affected yet and $\varepsilon$ is fixed.
According to Eq.~(\ref{eq:x1}),
\begin{equation}
x_1^{\prime}(r_s)\simeq 1+\frac{1}{\sqrt{\ln(r_s/b)}}.
\end{equation}

In order to discuss $x_m(r_s)$, we consider two different cases,
$qb/e\ll r_s\ll Qb/e$ and $b\ll r_s\ll qb/e$. When $qb/e\ll r_s\ll
Qb/e$, the chemical potential $\mu_c$ is still given by
Eq.~(\ref{eq:muc}). From Eq.~(\ref{eq:xo}),
\begin{equation}
x_m(r_s)=1+\frac{\ln(q/e)}{2\ln(r_s/b)}.
\end{equation}
Accordingly, the critical value of $q$ at which
$x_1^{\prime}(r_s)=x_m(r_s)$ is
\begin{equation}
q_c(r_s)=\exp\left[2\sqrt{\ln(r_s/b)}\right].
\end{equation}
It decreases with decreasing $r_s$. Therefore, for a system with
$q<q_c(\infty)$, $q$ can be larger than $q_c(r_s)$ at small $r_s$.
Correspondingly, at $x>1$, for small $r_s$, the phase of total
condensation is replaced by the phase coexistence of free
complexes and the single drop. We have a phase diagram shown in
Fig.~\ref{fig:phasehybrid}. Letting $q=q_c(r_s)$, we get the
critical value of $r_s$ at which this phase transition happens
\begin{equation}
r_0=b\exp\left(\frac{\ln^2(q/e)}{4}\right).\label{eq:r0}
\end{equation}
Notice that $r_0$ is much larger than $qb/e$.

In Fig.~\ref{fig:phasehybrid}, $x_1^{\prime}(r_s)>x_m(r_s)$ at
$r_s>r_0$, while $x_1^{\prime}(r_s)<x_m(r_s)$ at $r_s<r_0$
($x_1^{\prime}(r_s)$ curve at $r_s>r_0$ is not shown). By
definition of $r_0$, the curves $x_1^{\prime}(r_s)$ and $x_m(r_s)$
merge at $r_s=r_0$. The curve $x^{\prime}_s(r_s)$ is given by
Eq.~(\ref{eq:xs3}) for $r_s>r_0$ and Eq.~(\ref{eq:x1}) for
$r_s<r_0$. At $r_s=r_0$, these two expressions are equal to each
other. At $x>1$ side, the solid line at $L/r_s=L/r_0$ corresponds
to a first order phase transition. Notice that $r_0$ can be either
smaller or larger than $r_c$. Here for simplicity, only the former
case is shown in Fig.~\ref{fig:phasehybrid}.

When $b\ll r_s\ll qb/e$,
\begin{equation}
\mu_c=-\frac{qe}{Db}\ln(r_s/b),
\end{equation}
and $x_m(r_s)=3/2$~\cite{Nguyen-fraction}. In this case, we always
have $x_1^{\prime}(r_s)<x_m(r_s)$ as shown in
Figs.~\ref{fig:phase} and~\ref{fig:phasehybrid}.

Finally, in all cases discussed above, the value of $x_1^{\prime}$
is close to 1, i.e., $|n-n_i|\ll n_i$ in the condensation regime.
Therefore the approximation used in Sec.~\ref{sec:nosalt} that
$\mu_c$ is a constant is valid.

\section{The role of the translational entropy of polycations}\label{sec:entropy}
A major approximation in this paper is that the translational
entropy of PCs is negligible (we can always ignore PA's
translational entropy since it is much longer than PC). In this
section, we would like to discuss the validity of this
approximation and the role of the translational entropy.

First, let us estimate when this approximation is valid. Consider
PCs with concentration $p$ in the solution. The free energy due to
its translational entropy is $k_BT\ln(pv_0)$, where $v_0$ is the
normalizing volume. One the other hand, according to
Eq.~(\ref{eq:muc}), the Coulomb energy is in the order of
$-qe/Db\simeq -qk_BT/e$. They are equal at the critical value,
$p=\exp(-q/e)/v_0$. Therefore for a long PC with large $q$, we can
ignore PC's translational entropy even at exponentially small $p$.

If PC is very short, its translational entropy should be included.
DNA with short polyamines is a good example of such
systems~\cite{Olvera,Olvera2,Rouzina}. In this case, the phase
diagram gets another dimension, say, the concentration of PC, $p$.
The effect of PC entropy was discussed in detail in
Ref.~\cite{Nguyen-reentrant,Rui} in which the phase diagram is
drawn in a plane of two concentrations of oppositely charged
colloids at given $r_s$. Here we discuss the same effect in the
language of total charge ratio $x$ used in this paper in the
simple case where $x_1^{\prime}>x_m$ and $r_s\rightarrow\infty$.
For simplicity, we neglect the possibility of intra-molecule
disproportionation and the tadpole phase.

In this case, the free energy in Eq.~(\ref{eq:fyz5}) gets an
additional term due to the translational entropy of
PCs~\cite{Nguyen-reentrant}
\begin{eqnarray}
f(n,y)&=&y\left[\frac{(nq-Q)^2}{2C}+nE(n)\right]
+(1-y)[n_iE(n_i)+\varepsilon]\nonumber\\
&+&\left[\frac{xQ}{q}-yn-(1-y)n_i\right]
\ln\left[\left(p-\frac{ynq}{xQ}p-\frac{1-y}{x}p\right)
\frac{v_0}{e}\right],
\end{eqnarray}
where $e$ is the natural exponential. Here the expression in the
square bracket before the logarithm represents the number of free
PC in the solution, while the expression in the round bracket in
the logarithm represents their concentration.
\begin{figure}[ht]
\begin{center}
\includegraphics[width=0.7\textwidth]{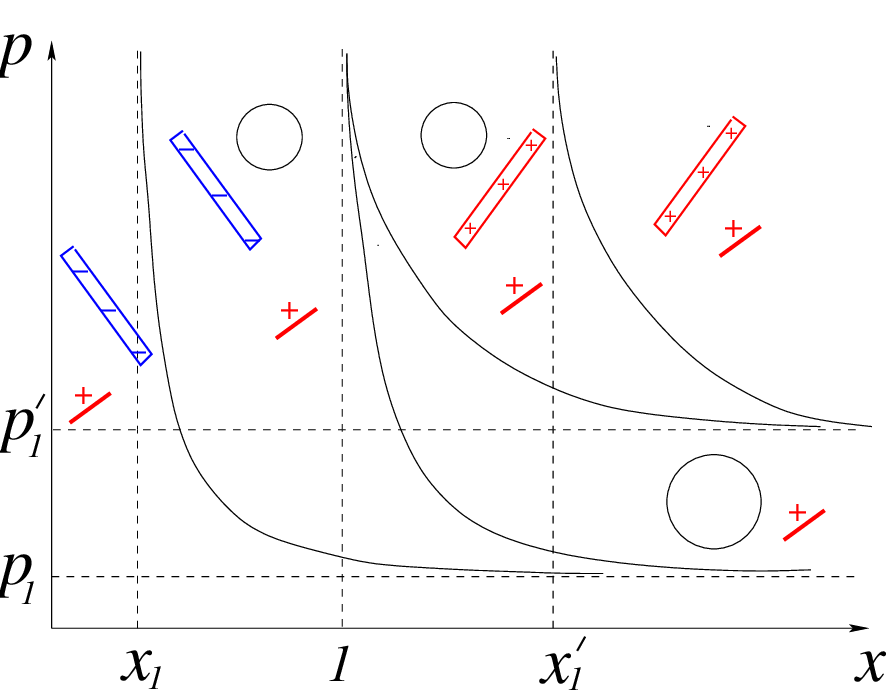}
\end{center}
\caption{Phase diagram in a plane of PC's concentration $p$ and
total charge ratio $x$ ($x_1^{\prime}>x_m$ and
$r_s\rightarrow\infty$). Symbols used are explained in
Fig.~\ref{fig:PE}. It shows how the phase of total condensation
replaces that of partial condensation with decreasing
$p$.}\label{fig:entropy}
\end{figure}

Now $n$ and $y$ are two independent variables. Taking $\partial
F/\partial n=0$ and $\partial F/\partial y=0$, we get
\begin{eqnarray}
\mu_c&=&\ln\left[\left(p-\frac{ynq}{xQ}p-\frac{1-y}{x}p\right)v_0\right]
-\frac{(nq-Q)q}{C},\label{eq:xcri}\\
\varepsilon&=&\frac{(nq-Q)^2}{2C}+(n-n_i)\left\{\mu_c-
\ln\left[\left(p-\frac{ynq}{xQ}p-\frac{1-y}{x}p\right)v_0\right]\right\}.
\label{eq:epsilon}
\end{eqnarray}
In Eq.~(\ref{eq:xcri}), eliminating $n$ by Eq.~(\ref{eq:epsilon}),
we can calculate the boundaries of the condensation regime by
setting $y=0$ and $y=1$. For $y=0$, we get two boundaries of a
total condensation phase,
\begin{eqnarray}
p\left(1-\frac{1}{x}\right)&=&p_1,\\
p\left(1-\frac{1}{x}\right)&=&p_1^{\prime}.
\end{eqnarray}
For $y=1$, we get two boundaries of the free complexes phase,
\begin{eqnarray}
p\left(1-\frac{x_1}{x}\right)&=&p_1,\\
p\left(1-\frac{x_1^{\prime}}{x}\right)&=&p_1^{\prime}.
\end{eqnarray}
Here
\begin{eqnarray}
p_1&=&\frac{1}{v_0}
\exp\left(\mu_c-\sqrt{\frac{2|\varepsilon|q^2}{C}}\right),\\
p_1^{\prime}&=&\frac{1}{v_0}
\exp\left(\mu_c+\sqrt{\frac{2|\varepsilon|q^2}{C}}\right).
\end{eqnarray}
Accordingly, as shown in the phase diagram Fig.~\ref{fig:entropy},
a regime of total condensation is sandwiched by two regimes of
partial condensation, which are further sandwiched by two regimes
of free complexes.

In Fig.~\ref{fig:entropy}, results of previous sections are
recovered in the limit $p\rightarrow\infty$. Due to the
translational entropy of PC, at finite $p$ all critical $x$ are
shifted to higher values. At the same time, a total condensation
region acquires a finite width even in the absence of monovalent
salt~\cite{Nguyen-reentrant}. In the limiting case where
$x\rightarrow\infty$, the concentration of PA is much smaller than
that of PC, and the entropy of PC is fixed, which offers a fixed
charging voltage to PA. As a result, all PA-PCs complexes are
either totally condensed or totally free. The partial condensation
regime disappears~\cite{Nguyen-reentrant}.

\section{Conclusion}\label{sec:conclusion}
In this paper, we discussed complexation and condensation of PA
with PC in a salty water solution. Using ideas of
disproportionation of PCs among complexes and inside complexes
(inter- and intra-complex disproportionations) we arrived at the
two phase diagrams in a plane of $x$ (ratio of total charges of PC
and PA) and $L/r_s$ (ratio of the length of PA, $L$, and the
Debye-H\"{u}ckel screening radius, $r_s$) shown in
Figs.~\ref{fig:phase} and~\ref{fig:phasehybrid}. In the case of
strongly charged PA and PC, we find that both two phase diagrams
are possible depending on the relative length of PC to PA.
Fig.~\ref{fig:phase} corresponds to a more generic case of
relatively long PC, while Fig.~\ref{fig:phasehybrid} to the case
of relatively short one. Our phase diagrams show how total
condensation is replaced by the partial one and then by phases of
stable complexes when $x$ moves away from $x=1$.

We discovered two new features of the phase diagrams. First, at
large screening radius they include a new phase of tadpoles and
corresponding phase coexistence. Second, we found that the phase
of the single drop formed at $x$ close to 1 widens with decreasing
$r_s$ as $1/r_s^{2}$.

Although we talked about strongly charged PA and PC one can also
consider phase diagram of weakly charged PA and PC and develop a
microscopic theory for it. In both cases, the qualitative picture
is the same since our discussion of the phase diagram is rather
general and independent of the microscopic mechanism of the
short-range attraction.

As mentioned in the introduction, the problem we solved in this
paper should be considered as an example of a more general problem
of the phase diagram of the solution of two oppositely charged
colloids. Another important system of this kind is a long PA with
many strongly charged positive spheres. When long double helix DNA
plays the role of PA, this system is a model for the natural
chromatin. Therefore, we call such system artificial
chromatin~\cite{Nguyen-reentrant}. Our phase diagrams with all new
features including tadpoles should be valid for artificial
chromatin as well.

There is another class of systems where only some of our
predictions are applicable. In the Ref.~\cite{Rui} we considered
solution of large negative spheres with positive spheres which are
smaller in both radius and charge. Complexation and condensation
of such spheres obey the phase diagrams similar to discussed
above. For example, screening by monovalent salt again leads to
$1/r_s^{2}$ expansion of the range of the single drop (total
condensation) phase. Another our prediction, the tadpole phase,
however, is not applicable to this case, because it is essentially
based on the polymer nature of PA.

In a case when the role of PA is played by DNA, one should
remember that the double helix DNA is so strongly charged that the
effect of the Manning condensation by monovalent counterions must
be included~\cite{Nguyen-reentrant}. Since DNA complexes with
positive macroions (PC, positive spheres or multivalent cations),
this Manning condensation can be weaker than free DNA due to
counterion releasing. The quantitative description of this effect
depends on the geometry of the positive macroions and the
microscopic structure of the complex in the
system~\cite{Nguyen-reentrant}. Generally speaking, this effect
leads to the renormalization of the bare charge of PA, $Q$. In
this case, total condensation still happens around $x=1$ but
renormalized charge enters in calculation of $x$. This means that
if on the other hand $x$ is evaluated using the bare charge of
DNA, all phase diagrams are centered around a smaller than 1 value
of $x$.

Our phase diagrams deal with equilibrium states of the system. But
not all of them can be achieved in experimental time scale due to
slow kinetics. Therefore it is not easy to directly compare our
theory to experiments. For instance, a phase of many condensed
particles with finite size is often found in experiments which
does not appear in our phase diagram~\cite{Kabanov,Budker,van
Zanten,Bloomfield}. We believe that this phase is not a real
equilibrium state, but the state frozen kinetically~\cite{TT,Ha}.
Thus, kinetics is extremely important for applications and we plan
to address it in the future.

\begin{acknowledgments}
The authors are grateful to V. Budker, A. Yu. Grosberg, and M.
Rubinstein for useful discussions. This work was supported by NSF
No. DMR-9985785 and DMI-0210844.
\end{acknowledgments}



\begin{thebibliography}{100}
\bibitem{ab} In this paper, for simplicity, we do not discriminate
between the size of the monomer and the distance between charges
in the polymer. This is a resonable approximation for strongly
charged polyelectrolytes ($b\simeq l_B=7 \AA$).
\end{thebibliography}
\end{document}